%Paper: hep-ph/9205247
%From: POMAROL@slacvm.slac.stanford.edu
%Date: Fri, 29 May 1992 20:00 -0800 (PST)

%macropackage=phyzzx
\def\SCIPP{\centerline {\it Santa Cruz Institute for Particle Physics}
  \centerline{\it University of California, Santa Cruz, CA 95064}}
\def\ie{{\it ie.}}

\def\calo{{\cal O}}

\def\calm{{\cal M}}

\def\hl{h^0}
\def\ha{A^0}
\def\hh{H^0}

\def\msusy{M_{SUSY}}

\def\sinb{{\rm sin}\beta}
\def\cosb{{\rm cos}\beta}

\def\hon{h^0_1}
\def\htw{h^0_2}
\Pubnum={SCIPP-92/19}
\date={May, 1992}
\pubtype{}
\titlepage
\vskip 4cm
\title{\bf Higgs sector CP violation
 in the Minimal Supersymmetric Model}
\author{Alex Pomarol}
\vskip .1in
\SCIPP
\vskip .2in

\vfill
\centerline{\bf Abstract}
We study the possibility that CP  is spontaneously broken
in the Minimal Supersymmetric Model when radiative corrections to the
Higgs potential are included. We show that this can only occur if
a light Higgs boson exists. Considering the recent ALEPH Higgs search, we
exclude most of the parameter space of the model.
The possibility of explicit CP violation in the model
is also briefly discussed.
\vskip .1in

\vfill
\endpage
It has been known for a long time that
 when supersymmetry (SUSY) is imposed
on the two Higgs doublet model (THDM), tree-level
flavor changing neutral
\REF\mssm{For a comprehensive review and a guide to the literature, see
J.F. Gunion, G.L. Kane, H.E. Haber and S. Dawson,
 {\it The Higgs Hunter's Guide} (Addison-Wesley Publishing Company,
  Reading, MA, 1990).}
  currents and CP violation are simultaneously avoided in the
 Higgs sector\refmark\mssm .
 Nevertheless, since
SUSY must be softly broken, new terms in
the Higgs potential can be induced by radiative corrections and CP
non-conserving effects could show up in the Higgs sector.
CP may be broken in two different ways: explicitly and
spontaneously. In the first case,
 CP violation derives  from complex scalar
self-couplings induced by radiative corrections by sectors of the
 theory which violate CP.
In the second case, a
 relative phase between the vacuum expectation values (VEVs)
 of the two Higgs doublets arises
 which spontaneously breaks the CP
\REF\lee{T.D. Lee, Phys. Rev. {\bf D8} (1973) 1226; Phys. Rep. {\bf 9}
(1974) 143.}
 symmetry\refmark\lee .

The purpose of this paper is to study the possibility that CP violation
 appears in one of these ways in the Higgs sector of the minimal
supersymmetric model (MSSM).
  In such a  case
the CP-even and CP-odd neutral scalars would mix with each other
 giving rise
\REF\con{See e.g.,
C. Jarlskog, editor, {\it CP violation} (World Scientific,
 Singapore, 1989).}
to important phenomenological consequences\refmark\con .
 \REF\scpv{N. Maekawa, Preprint KUNS 1124 (1992).}
It was recently pointed out\refmark\scpv\ that spontaneous
 CP violation (SCPV)
 in the MSSM can occur.
However, it is known that in order that radiative corrections can cause
\REF\lig{H. Georgi and S.L. Glashow, Phys. Rev. {\bf D6} (1972) 2977;
\hfil\break
 S. Coleman and E. Weinberg, Phys. Rev. {\bf D7} (1973) 1888.}
\REF\gp{H. Georgi and A. Pais, Phys. Rev. {\bf D10} (1974) 1246;
{\bf D16} (1977) 3520.}
a spontaneous broken vacuum, a light scalar is
 required\refmark{\lig,\gp}.
Therefore, an analysis of the physical spectrum, not carried
 out in ref. [\scpv], is necessary in order to determine  the viability
 of this model.

Let $\Phi_1$ and $\Phi_2$ denote two Higgs doublets with hypercharges
$Y=1$.
The most general renormalizable $SU(2)_L\times U(1)_Y$ gauge invariant
 two Higgs doublet potential is given by
$$\eqalign{V(\Phi_1,\Phi_2)=&\ m^2_{1}\Phi_1^\dagger\Phi_1
+m^2_{2}\Phi_2^\dagger\Phi_2-(m^2_{12}\Phi_1^\dagger\Phi_2+h.c.)\cr
+&\ \lambda_1(\Phi_1^\dagger\Phi_1)^2+
\lambda_2(\Phi_2^\dagger\Phi_2)^2+\lambda_3(\Phi_1^\dagger\Phi_1)(
\Phi_2^\dagger\Phi_2)
+\lambda_4(\Phi_1^\dagger\Phi_2)(\Phi_2^\dagger\Phi_1)\cr
+&\ \coeff{1}{2}\left[\lambda_5(\Phi_1^\dagger\Phi_2)^2+h.c.\right]
+\coeff{1}{2}\left[\Phi_1^\dagger\Phi_2\{  \lambda_6(\Phi_1^
\dagger\Phi_1)+
\lambda_7(\Phi_2^\dagger\Phi_2)\}+h.c.\right]  \, ,}\eqn\potential$$
where by hermiticity only $m^2_{12}$, $\lambda_5$, $\lambda_6$ and
 $\lambda_7$ can be complex. Let us first consider the case where
  these parameters are real, \ie\ CP is not explicitly violated.
   After spontaneous symmetry breaking, the VEVs of the
   neutral components of the
   Higgs doublets are given by
$$<\phi_1^0>=v_1\ ,\ \ \ \ \ \ <\phi_2^0>=v_2e^{i\xi}\, .$$
In order to have SCPV, \ie\ $\xi\not=\coeff{n\pi}{2}\ (n\in {\cal Z})$,
 we need
$$\lambda_5>0\ ,\eqn\cpvco$$
$$\left|\coeff{2m^2_{12}-\lambda_6v_1^2-\lambda _7v_2^2}
{4\lambda_5v_1v_2}\right|<1\, .\eqn\cpvc$$
In this case, at the minimum of the potential,
$$\cos\xi=
\coeff{2m^2_{12}-\lambda_6v_1^2-\lambda _7v_2^2}
{4\lambda_5v_1v_2}\, .$$
When SUSY is imposed on the two Higgs doublet potential
 we have\refmark\mssm\ ,
$$\lambda_1=\lambda_2=\coeff{1}{8}(g^2+g^{\prime 2})\, ,\ \ \
\lambda_3=\coeff{1}{4}(g^2-g^{\prime 2})\, ,\ \ \ \lambda_4=-\coeff{1}{2}
g^2\, ,\ \ \ \lambda_5=\lambda_6=\lambda_7=0\, .$$
Thus, eq. \cpvc\ does not hold and
$\xi$ must be $0$ or $\pi$.
When radiative corrections are considered, new terms in the Higgs
potential are induced.
In the limit where the SUSY scale is large, $\msusy \gg m_W$, only terms
 of dimension less than or equal to 4 are not suppressed by inverse
  powers of $\msusy$. In this limit,
   the effective low-energy Higgs potential
of the MSSM is given by eq. \potential.

In order to know whether eqs. \cpvco\ and
\cpvc\ hold, we must calculate the induced
$\lambda_5$ parameter.
The   $\lambda_6$ and $\lambda_7$ parameters   are in fact not relevant
  because $m^2_{12}$ is a free parameter.
The only contribution that generates a positive $\lambda_5$ comes from
\FIG\fig{One-loop Feynman diagram contributing to $\lambda_5$.}
diagrams involving loops of charginos and neutralinos (fig. \fig ).
Squark and Higgs loops give a negative contribution to $\lambda_5$
but they can be neglected in the case of
 small $\tilde q_R-\tilde q_L$ mixing and
 small $m^2_{12}$ respectively. Quarks and gauge bosons do not
contribute.
In the limit of equal mass charginos and neutralinos, we find
$$\lambda_5=\coeff{g^4}{32\pi^2}\sim 5\cdot  10^{-4}.$$
Therefore, we see from eq. \cpvc\ that,
in order that SCPV occur, the tree level parameter
$m^2_{12}$ must be of \calo($\lambda_5 v_1v_2) \sim (3\ {\rm GeV})^2\ .$
This seems to contradict the Georgi--Pais theorem\refmark\gp\ which
says that SCPV can only be generated by radiative corrections
when a tree-level massless scalar field,
 other than the Goldstone boson, exists
\foot{We must have  $m^2_{12}=0$  in order to have a
massless Higgs boson ($A^0$) at tree-level.}.
 Notice, however,
that this theorem is strictly true only for
first order corrections to the effective potential.
When two-loop corrections are considered, it is easy to see that
the scalar can have a tree-level mass whose magnitude is
of one-loop order\refmark\gp .
 Of course,
 the theorem can only be applied when the true minimum is close to the
tree-level minimum.

 To analyze the physical spectrum, let us make the following
rotation
$$\eqalign{\Phi^\prime_1=&\ \cosb\ \Phi_1+\sinb\ e^{-i\xi}\Phi_2=
\left(\matrix{G^+\cr v+\coeff{1}{\sqrt{2}}\left(\hl +iG^0 \right)}
\right)\, ,\cr
\Phi^\prime_2=&-\sinb\ \Phi_1+\cosb\ e^{-i\xi}\Phi_2=
\left(\matrix{H^+\cr \coeff{1}{\sqrt{2}}\left(\hh +i\ha\right)}
\right)\, .}$$
where $\tan\beta=v_2/v_1$, $v=\sqrt{v^2_1+v^2_2}$,
$G^+$ and  $G^0$ are the goldstone bosons, $\hl$ and  $\hh$ are
\REF\men{A. M\'endez and A. Pomarol, Phys. Lett.
  {\bf B272} (1991) 313.}
CP-even fields  and $\ha$ is  a CP-odd field\refmark\men.
The three physical neutral Higgs boson mass eigenstates are mixtures
    of $\hl$, $\hh$ and $\ha$.
The relevant elements of the neutral scalar mass matrix are given by
$$\eqalign{\calm^2_{\hl\ha}=& -2m^2_{12}\sin\xi\, ,\cr
\calm^2_{\hh\ha}=&\ \left[\lambda_5(v^2_2-v^2_1)\cos\xi+
(\lambda_6-\lambda_7)v_1v_2\right]\sin\xi\, ,\cr
\calm^2_{\ha\ha} =&\ 2\lambda_5(v^2_1+v^2_2)\sin^2\xi\, .}\eqn\mass$$
It is clear from  eq. \mass\ that there is a light Higgs boson
for any value of $\xi$
and $\tan\beta$.

Let us first consider the case where the other neutral Higgs bosons
 are much heavier. These
  will be predominantly CP-even states with a small
  admixture of
$\ha$. In this case our model will be similar
 to the MSSM without CP violation and with a light $A^0$:
$$m^2_{\ha}\simeq \calm^2_{\ha\ha} \lsim (6\ {\rm GeV})^2\, .$$
\REF\lep{D. Decamp et al. (ALEPH Collaboration),
 Phys. Lett. {\bf B265} (1991) 475.}
Since the recent limit from ALEPH Collaboration\refmark\lep\ implies
 a lower
 bound of $20$ GeV
for the CP-odd scalar mass, this possibility is ruled out\foot{
The ALEPH limit is only valid in the region $\tan\beta >1$.
For $\tan\beta <1$   there exists a region of the MSSM
parameter space in which a light $A^0$ is not excluded by ALEPH.
However, regions of parameter space where $\tan\beta<1$ are strongly
\REF\sugra{See e.g., G.F. Giudice and G. Ridolfi, Z. Phys. {\bf C41}
(1988) 447;\hfill\break
M. Olechowski and S. Pokorski, Phys. Lett. {\bf B214} (1988) 393;\hfill
\break
M. Drees and M.M. Nojiri, Nucl. Phys. {\bf B369} (1992) 54.}
disfavored in low-energy supersymmetric models\refmark\sugra.}.

A second possibility  is that the mass of one of the CP-even scalars
 is also small and mixes substantially with  the
$\ha$. In this case, the ALEPH data must be
carefully examined to determine if this possibility is excluded.
In particular, the lower scalar mass limits from ALEPH are not
 valid in a CP violating THDM.
To see why this is so, let us denote by $\hon$
 and $\htw$  the two lightest Higgs bosons, and by
$\coeff{g(p_1+p_2)^\mu}{2\cos\theta_W}\Theta_{\hon\htw Z}$ and
$\coeff{igm_Z}{\cos\theta_W}g^{\mu\nu}
\Theta_{h^0_i ZZ}$ the Feynman rules for
the tree level $\hon\htw Z$ and $h^0_i ZZ$ ($i=1,2$) couplings
respectively.
For a CP conserving Higgs sector ($\htw\equiv A^0 $),
$$\Theta_{\hon ZZ}^2+\Theta_{\hon\htw Z}^2=1\, .\eqn\sr$$
This relation, which plays a crucial role in inferring lower
 mass limits
 for the Higgs bosons, need not be satisfied when CP is violated
  in the Higgs sector. Nevertheless, a
  sum rule similar to eq.
  \sr\ can be also derived for a CP violating THDM.
It is given by\refmark\men
$$\Theta_{\hon ZZ}^2+\Theta_{\htw ZZ}^2+\Theta_{\hon\htw Z}^2
=1\, .\eqn\ort$$
On the other hand, assuming that $m_{\hon}\simeq m_{\htw}\lsim 20\
 {\rm GeV}$, the ALEPH Higgs search\refmark\lep\ implies
 the  following limits on the $\Theta$'s:
$$\eqalign{\Theta_{h^0_i ZZ}^2\lsim & 0.1\, ,\cr
\Theta_{\hon\htw Z}^2\lsim & 0.7\, .}\eqn\cons$$
Combining eq. \ort\ and eq. \cons\ we can  rule out the possibility of
two light Higgs bosons of indefinite CP.
 If the $\hon$ and
$\htw$ were light enough, they would decay outside the detector and
the constraints of eq. \cons\ could not be deduced.
 Nevertheless, since any
contribution to the $Z$ width from non-standard processes is limited
\REF\wit{D. Decamp et al. (ALEPH Collaboration),
 Preprint CERN-PPE/91-105 (1991).}
to less than $0.26\ \Gamma_{\nu\bar\nu}$\refmark\wit,
 bounds on the $\hon\htw Z$ and
\REF\lept{D. Decamp et al. (ALEPH Collaboration),
 Phys. Lett. {\bf B262} (1991) 139.}
$h^0_i ZZ$ couplings can also be inferred\refmark{\lep,\lept},
which turn out to be  in
 contradiction with eq. \ort.

Finally, let us briefly
consider the case when other sectors of the theory
 violate CP. In that case, the induced couplings $\lambda_5$, $\lambda_6$
 and $\lambda_7$ can be complex\foot{$m^2_{12}$ can be made real by
  a redefinition of the Higgs doublets.} and we have a CP violating
Higgs sector even for real VEVs. In supersymmetric
 theories there
 are a number of new sources of CP violation from the various
 supersymmetric sectors.
 Nevertheless, experimental limits on the neutron
 electric dipole moment require any such
 CP violating phases, $\varphi$,
\REF\ph{J. Ellis, S. Ferrara and D.V. Nanopoulos, Phys. Lett. {\bf B114}
(1982) 231;\hfil\break
W. Buchm\"uller and D. Wyler, Phys. Lett. {\bf B121} (1983) 321;\hfil
\break
J. Polchinski and M. B. Wise, Phys. Lett. {\bf B125} (1983) 393;\hfil
\break
F. del Aguila, M.B. Gavela, J.A. Grifols and A. M\'endez, Phys. Lett.
{\bf B126} (1983) 71;\hfill\break
R. Arnowitt, J.L. Lopez and D.V. Nanopoulos, Phys. Rev. {\bf D42} (1990)
2423;\hfill\break
R. Arnowitt, M.J. Duff and K.S. Stelle, Phys. Rev {\bf D43} (1991) 3085.}
 to be less than  $10^{-2}$\refmark\ph . As a result, explicit CP
 violation effects in the
Higgs potential will be of order
$$\lambda_{5,6,7}\cdot\varphi\sim 10^{-5}\, .$$
These effects are too small to have any significant
phenomenological implications.

Summarizing, we have seen that the MSSM with SCPV
require  the existence of a  Higgs boson  with a  mass of the order
      of a few GeV.
 Based on the recent ALEPH Higgs search\refmark\lep,
 we have seen that this model
  is easily ruled out (except perhaps
  for a small disfavored region of parameter
space where $\tan\beta<1$). Although explicit CP violation
in the Higgs sector is in principle possible, it turns out to be too
 small to be phenomenologicaly relevant.
\vskip .5cm
\centerline{\bf Acknowledgements}

I would like to thank Howard  Haber for helpful conversations and
for a critical reading of the manuscript.
This work was supported by a fellowship of MEC (Spain).
\endpage
\refout
\endpage
\end